\documentclass[twocolumn,prb,showpacs,preprintnumbers,amsmath,amssymb,citeautoscript]{revtex4-2}

\usepackage{graphicx}
\usepackage{dcolumn}
\usepackage{bm}
\usepackage{epstopdf}
\usepackage{subcaption}
\usepackage{color,hyperref}
\definecolor{blue}{rgb}{0.0,0.0,1}
\hypersetup{colorlinks,breaklinks,
            linkcolor=blue,urlcolor=blue,
            anchorcolor=blue,citecolor=blue}
\usepackage{caption}
\begin{document}

\title{Bond-Strength-Based Understanding of Oxygen Vacancy Migration Barriers in Rutile Oxides}
\author{Inseo Kim}
\email{rmfosem753@gmail.com} 
\author{Minseok Choi}
\email{minseok.choi@inha.ac.kr} 
\affiliation{Department of Physics and Institute of Quantum Science, Inha University, Incheon 22212, Korea}
\date{\today}
\begin{abstract}

We carry out bond-strength based analysis for the migration barrier ($E_{\rm B}$) of oxygen vacancies in rutile-type 3$d$ transition-metal 
dioxides by combining density-functional theory (DFT) and the bond-valence model. The covalent and ionic contributions to chemical bonding
are explicitly decomposed and quantified by the sum of the integrated crystal orbital Hamilton population ($S_c$) and the Madelung energy ($S_i$), respectively. 
Both $S_c$ and $S_i$ exhibit strong correlations with the $E_{\rm B}$ from DFT ($E_{\rm B}^{\rm DFT}$), and their average $\bar{S}$ provides a reasonable estimate 
of $E_{\rm B}^{\rm DFT}$ across the oxide series. Inspired by the bond-valence model, two parameters are extracted by fitting to a large dataset of 3$d$ transition-metal 
dioxides. Our results show that using these parameters, $E_{\rm B}$ of oxygen vacancies can be efficiently estimated. 
\end{abstract}
\maketitle

\section{Introduction}

Ion migration barrier ($E_{\rm B}$) and pathway are crucial to address key functionality of materials such as ionic conductivity, memristive switching, and redox kinetics ~\cite{yildiz2014stretching,waser2007nanoionics}. Density-functional theory (DFT) calculation has successfully addressed them~\cite{defect_freysoldt2014first,defect_nitrides,defect_oba2011point}, typically by considering the motion of one atom in conjunction with the movement of its neighboring atoms. A prototypical example is oxygen vacancies ($V_{\rm O}$) in oxide. The $V_{\rm O}$ migration has been understood by examining the movement of a certain oxygen atom (the counterpart O atom of $V_{\rm O}$ that replaces its position with $V_{\rm O}$) and its neighboring cations and anions. The plausible $E_{\rm B}$ and migration pathway are determined by investigating the $V_{\rm O}$-induced properties such as defect levels, local atomic features, and configuration-energy diagrams~\cite{limpijumnong2004diffusivity,mayeshiba2015strain,mayeshiba2016factors,alaydrus2021mechanistic,zhao2016oxygen}.


Accurate assessment and prediction of $E_{\rm B}$ using DFT-based approaches remains a challenge, mostly due to the demanding huge computational cost. In particular, the climbing image nudged elastic band (cNEB) method, one of the most sophisticated methods to find $E_{\rm B}$, requires multiple constrained DFT calculations along the migration pathway, significantly increasing the computational cost~\cite{neb}. There also exists the uncertainty partly caused by the choice of exchange--correlation functional~\cite{devi2022effect,santana2017diffusion}. These can hinder accelerating materials design (e.g., novel electrolyte), for instance, through high-throughput screening, and improving the functionality of existing materials (e.g., high ionic conductivity). 

$E_{\rm B}$ should be related to the bonding nature between a migrating ion and its environment. Because the ion migration necessarily involves bond breaking and reformation processes at the saddle point, where the local bonding environment differs substantially from that of the equilibrium site ~\cite{yang2023oxide}. It would closely correlate with bond strength associated with the migrating ion, and it is thus natural to utilize the method enabling the quantification of bond strength. 

The crystal orbital Hamilton population (COHP) method allows us to investigate chemical bonding in terms of energy-resolved bond interactions between atoms by considering overlapping electronic states~\cite{cohp1} and provides  the so-called integrated COHP (ICOHP) that is a quantitative measure of covalent bond strength. Several studies have suggested that the formation energy of $V_{\rm O}$ correlates with the ICOHP. Fung {\it et al}. showed that the absolute values of the ICOHP and the formation energy of $V_{\rm O}$ in perovskite SrTMO$_3$ (TM = transition metal) are linearly and monotonically correlated~\cite{fung2018new}. A similar relationship has also been reported for strained Ti-based oxides~\cite{kim2022first}. 

An earlier theoretical framework, the bond-valence model (BVM), is also known to enable the description of chemical bonding in terms of overlapping atomic orbitals. The BVM provides an expression for the oxidation state of atom $i$ ($V_i$) to quantify bond strength, given by the following equation,
\begin{eqnarray} \label{eq1}
	V_i= \sum_{j}s_{ij}= \sum_{j}exp\left ( \frac{r_{0}-r_{ij}}{b}\right),
\end{eqnarray}
where $s_{ij}$ and $r_{ij}$ are the bond valence and the bond length between atoms $i$ and $j$, respectively. The parameters $r_{0}$ and $b$ represent the nominal bond length and an universal constant, respectively, obtained by fitting to experimental data~\cite{brown1985bond,brown2009recent}. Since its introduction, the BVM has been continuously refined through revisions of originally proposed bond parameter values and through its integration with other theoretical approaches. Recently, Adams and Rao developed a force-field method by combining the BVM with a Morse-type interaction potential~\cite{adams2009transport,adams2011high}. They developed an approach, enabling the calculation of ion migration pathways and energies with improved accuracy, which is closely related to the bond strength between the migrating ion and its surrounding atoms.

To our knowledge, however, no studies have exploited the explicit correlations between the ICOHP and $E_{\rm B}$ of $V_{\rm O}$ and have reported the relationship between ICOHP and BVM. 

In this work, our aim is to investigate the $E_{\rm B}$  of $V_{\rm O}$ based on the bond strength associated with the vacancy. We perform DFT-cNEB calculation to obtain $E_{\rm B}$ of $V_{\rm O}$ in rutile-type 3$d$--TMO$_2$ in which $V_{\rm O}$ play a decisive role, as they serve, for instance, as mobile ionic species whose drift under an applied electric field underpins resistive switching mechanisms~\cite{waser2007nanoionics}. Then, the $E_{\rm B}$ from DFT-cNEB are analyzed by explicitly decomposing bond strength into covalent and ionic contributions. 
Two ICOHP-based BVM parameters are extracted from the DFT-ICOHP results. Finally, an estimation of $E_{\rm B}$ based on the parameters is attempted and discussed.

\section{Computational details}
\label{sec:method}

First-principles calculations were performed using the projector augmented-wave method~\cite{paw} and the generalized gradient approximation (GGA) functional of Perdew-Burke-Ernzerhof scheme implemented in the {\sc vasp} code ~\cite{vasp}. For bulk calculation, the energy difference of $10^{-8}$ eV was a criterion for the electronic optimization, and the atom coordinates were optimized until the Hellmann-Feynman force was less than 0.001 eV/{\AA}. An energy cutoff of 700 eV was used, and the six atom unitcell and 11 $\times$ 11 $\times$ 17 $k$-point grid were used for the integration of the Brillouin zone. For simulating $V_{\rm O}$, the 2 $\times$ 2 $\times$ 3 supercells containing 72 atoms and 5 $\times$ 5 $\times$ 5 $k$-point grid were used. Atomic relaxation was conducted until the Hellmann-Feynman force was less than 0.02 eV/{\AA}.

The DFT-based $E_{\rm B}$ of $V_{\rm O}$ ($E_{\rm B}^{\rm DFT}$) and the migration pathway were obtained using the cNEB method~\cite{neb}. O$^*$ is set to be an oxygen that becomes $V_{\rm O}$, i.e., the atom removed from the supercells in the defect calculations. The migration of $V_{\rm O}$ was examined by displacing a neighboring O atom, which is crystallographically equivalent to O$^*$, toward the site of O$^*$ along the migration pathway. This equivalence ensures that the local bonding environment of the displacing atom is identical to that of O$^*$, such that the bond-strength quantities evaluated at O$^*$ are directly representative of those governing the migration process.





The ICOHP and the Madelung energy ($E_{\rm M}$) were obtained using the {\sc lobster} code \cite{cohp1,cohp2,cobi}. The ICOHP was obtained by integrating up to the Fermi level; negative values (i.e., --ICOHP) provide the covalent bond strength of chemical bonds in units of energy. Larger (smaller) negative ICOHP indicates stronger (weaker) covalent bond strength. $E_M$ was evaluated based on the partial atomic charges obtained through the Mulliken and L\"owdin population analyses. The oxidation states of atoms in the optimized structures were determined using the BVM, as implemented via the BVAnalyzer module in the {\sc pymatgen} package ~\cite{pymatgen}.  

The covalency of certain solid system was quantified using the integrated crystal orbital bond index (ICOBI)~\cite{cobi} that is also employed in the {\sc lobster} code; smaller ICOBI indicates a more ionic bonding character. 

\section{Results and discussion}
\label{sec:resdis}

\subsection{Bond strength}
\label{sec:bond}

\begin{figure}[]
\includegraphics[width=1\columnwidth]{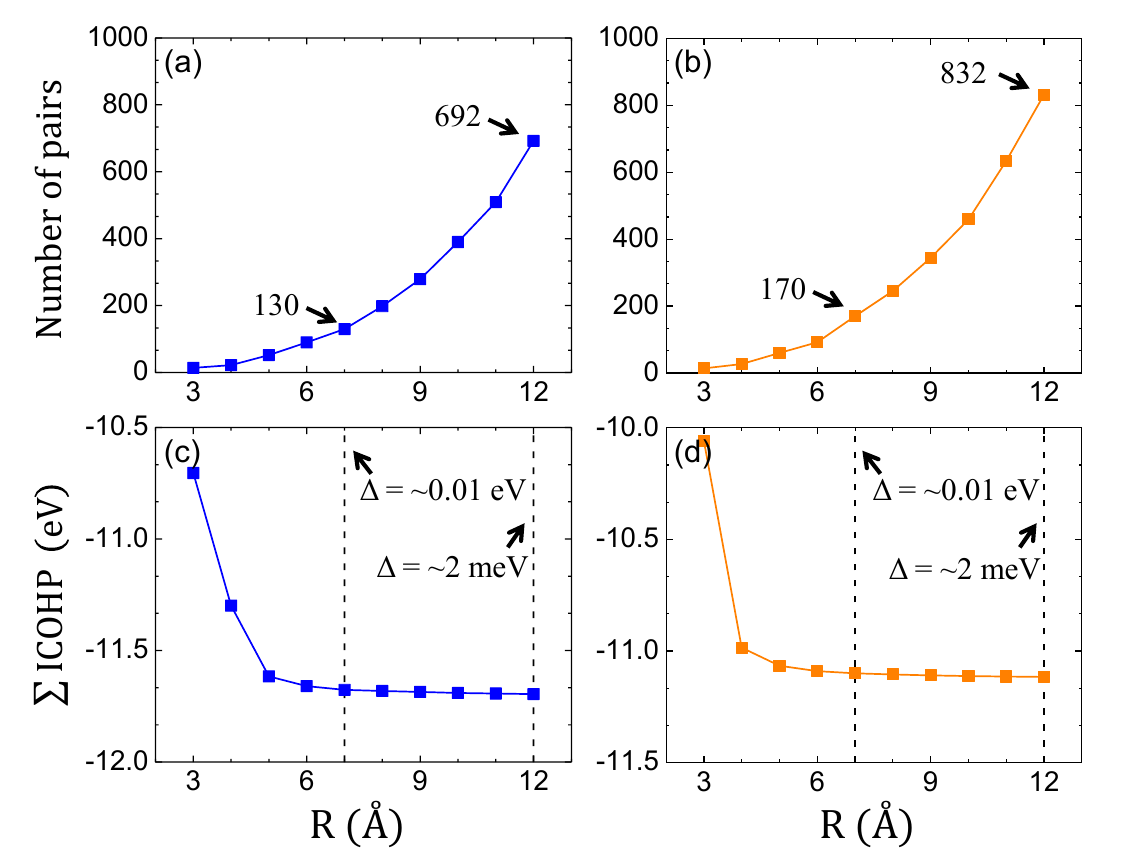}\\
\caption{ \label{FIG_bond} Number of $X$--O$^*$ pairs and $\sum$ICOHP as a function of $R$ (the radius of a shell centered at O$^*$). (a), (c) TiO$_2$ and (b), (d) FeO$_2$. }
\end{figure}


\begin{figure*}[]
	\includegraphics[width=2\columnwidth]{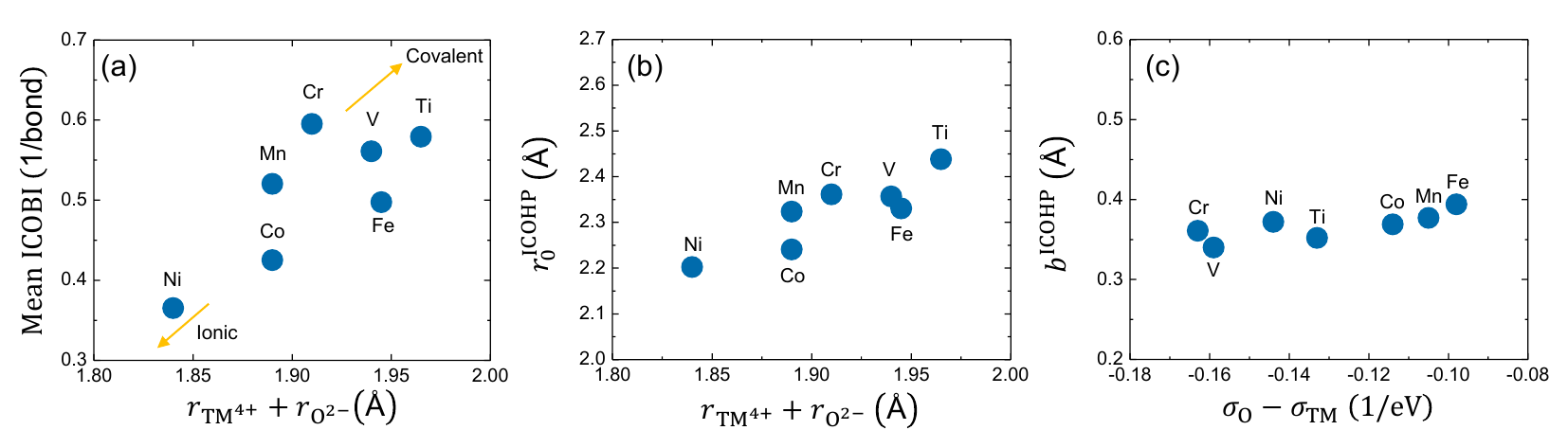}\\
	\caption{ \label{FIG_fit_2} (a) Mean ICOBI of six TM--O in TMO$_2$ as a function of the sum of the ionic radii of TM$^{4+}$ and O$^{2-}$. (b) $r_0^{\rm ICOHP}$ as a function of the sum of TM$^{4+}$ and O$^{2-}$ ionic radii. (c) Relationship between $b^{\rm ICOHP}$ and the softness difference between O and the TM element ($\sigma_{\rm O} - \sigma_{\rm TM}$)~\cite{hardness1}. The ionic radii of each element are taken into account by considering the coordination number and oxidation states~\cite{shannon}.}
\end{figure*}


Firstly, we examine {\it practical} covalent bond strength in TMO$_2$. The TM--O$^*$ bonds mainly contribute to the ICOHP, but the contributions from interatomic interactions with the rest of atoms in the simulation cell (X--O$^*$, X = TM, O) may not be excluded. 
In the BVM, it was reported that restricting bond-strength contributions to the first coordination shell alone systematically underestimates the total valence sum, as more than 11\% of the Li-O valence sum originated from interactions beyond the nearest-neighbor shell in Li-O compounds~\cite{Adams:br0103}. Analogous to this, the cumulative ICOHP was evaluated as a function of the interaction range $R$ (defined as a radius of shell centered at O$^*$) to determine an appropriate cutoff. 


Figures~\ref{FIG_bond} (a) and (b) show that the number of $X$--O$^*$ pairs within the shell with a radius $R$. 
The number of $X$--O$^*$ pairs dramatically increases when $R$ is more extended. 
Figures~\ref{FIG_bond} (c) and (d) show The sum of ICOHP ($\sum$ICOHP) to quantify the contribution from the $X$--O$^*$ pairs. $\sum$ICOHP converges within 0.01 eV for $R$$\sim$7 {\AA} and 2 meV for $R$$\sim$12 {\AA}. This seems to be connected to that cutoff radii in the range of 6--8 \AA\ in the BVM to ensure that weak interactions from extended coordination shells are not overlooked~\cite{Adams:br0103}. We therefore adopt $R$ = 7 \AA~ as the cutoff for evaluating covalent bond strength, ensuring that the dominant bond-strength contributions are captured without incurring unnecessary computational overhead from interactions of negligible magnitude.


The {\it practical} covalent bond strength ($S_{c}$) can then be defined as the $\sum$ICOHP using the following equation, 
\begin{eqnarray} \label{eq2}
S_{c}= \sum_{r \leq 7 \mathring{\!A}}\sum_{X}\mathrm{ICOHP}_{X-\rm O^*}(r),
\end{eqnarray}
for containing the $X$--O$^*$ pairs in the shell with $R = 7$ \AA.

In addition to $S_{c}$, primarily capturing covalent bonding characteristics, $E_M$ is also evaluated for ionic contributions, as follows:
\begin{eqnarray} \label{eq4}
	\textit E_\textit M=\frac 1 2 \sum_{i} q_iv_i.
\end{eqnarray}
where ${q_i}$ and ${v_i}$ are the charge and the electrostatic site potential of ion ${i}$, respectively. Then, the ionic bond strength (${S_i}$) is defined as
\begin{eqnarray} \label{eq5}
	S_{i}= \frac 1 2 \textit q [\rm {O^*}] \textit v [\rm {O^*}],
\end{eqnarray}
The evaluation of the $E_M$ is sensitive to the choice of partial atomic charges. To account for the dependence on the charge partitioning scheme, the Mulliken and L\"owdin population analyses were utilized and compared with each other. The former and the latter are denoted as $S_{i}^{M}$ and $S_{i}^{L}$, respectively.

The bond strength quantities $S_{c}$ and $S_{i}$ were evaluated using the relaxed structures of each image along the migrating pathway of $V_{\rm O}$ in the DFT-cNEB calculations, and the values of $S_{c}$ and $S_{i}$ at the initial state were set as reference.  

\begin{figure}[b]
    	\includegraphics[width= 8 cm]{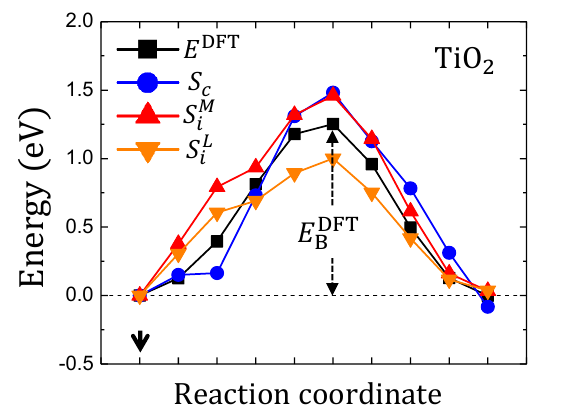}\\
	\caption{ \label{FIG_barrier} Energy profiles of migrating $V_{\rm O}$ in TiO$_2$ with $E^{\rm DFT}$, $S_{c}$, $S_{i}^{M}$, and $S_{i}^{L}$. All the values at the initial position (black arrow) are set as a reference. $E_{\rm B}^{\rm DFT}$ denotes the migration barrier obtained by using explicit DFT calculation. L and M in superscript stand for the L\"owdin and Mulliken approaches, respectively.
    }
    
\end{figure}

\begin{table*}[]
	\caption{\label{barrier1} Calculated $E_{\rm B}^{\rm DFT}$ and bond strength quantities; $S_{\rm c, max}$, $S_{i,\rm {max}}^{L}$, $S_{i,\rm {max}}^{M}$, $\bar{S}^{L}$, $\bar{S}^{M}$, $\tilde{S}^{L}$, and $\tilde{S}^{M}$ (see text). $L$ and $M$ in superscript stand for the L\"owdin and the Mulliken approaches, respectively. All the values are in unit of eV.}
	\begin{ruledtabular}
		\begin{tabular}{cccccccccc}
			& $S_{\rm c, max}$ & $S_{i,\rm {max}}^{L}$ & $S_{i,\rm {max}}^{M}$ & $\bar{S}^{L}$ & $\bar{S}^{M}$ & $\tilde{S}^{L}$ & $\tilde{S}^{M}$&  $E_{\rm B}^{\rm DFT}$ \\
			\hline
			TiO$_2$ & 1.48 & 1.00 & 1.46 & 1.24 & 1.47 & 1.28 & 1.47 & 1.25 \\
            VO$_2$  & 1.04 & 1.33 & 1.92 & 1.19 & 1.48 & 1.17 & 1.43 & 1.02 \\
            CrO$_2$ & 1.92 & 0.57 & 0.84 & 1.24 & 1.38 & 1.37 & 1.48 & 1.17 \\
            MnO$_2$ & 1.32 & 1.11 & 1.55 & 1.22 & 1.44 & 1.22 & 1.43 & 1.26 \\
            FeO$_2$ & 1.85 & 0.70 & 1.12 & 1.27 & 1.49 & 1.27 & 1.48 & 1.42 \\
            CoO$_2$ & 1.90 & 0.81 & 1.26 & 1.36 & 1.58 & 1.27 & 1.53 & 1.63 \\
            NiO$_2$ & 1.81 & 1.16 & 1.48 & 1.49 & 1.65 & 1.40 & 1.60 & 1.51 \\
            CuO$_2$ & 1.21 & 1.50 & 1.75 & 1.35 & 1.48 & 1.42 & 1.61 & 0.88 \\
          
		\end{tabular}
	\end{ruledtabular}
\end{table*}

According to textbook knowledge, the relative importance of covalent and ionic interactions varies substantially among different materials, reflecting variations in electronic structure, orbital hybridization, and charge transfer. The assignment of equal weights to $S_{c}$ and $S_{i}$ therefore implicitly assumes a uniform balance between covalency and ionicity across all systems. Such an assumption may not be physically appropriate, particularly for materials in which one bonding character is strongly dominant.


Accordingly, we attempt to introduce such material-dependent weighting by using the ICOBI. The ICOBI values of the six TM--O interactions were averaged to obtain a mean ICOBI for each compound. Figure~\ref{FIG_fit_2}(a) shows the mean ICOBI of TM--O$^*$ bonds as a function of the sum of ionic radii. The mean ICOBI is seen to correlate with 3$d$ electron count of TM. The values decrease as the 3$d$ electron count of TM increases. The mean ICOBI value of the Ti$^{4+}$(3$d^0$)--O$^{2-}$ bond is 0.579. The smallest value is found for Ni$^{4+}$(3$d^6$)--O$^{2-}$ bond (0.365), and the largest value is observed for the Cr$^{4+}$(3$d^1$)--O$^{2-}$ bond, reaching 0.595 that is almost twice of that of Ni$^{4+}$--O$^{2-}$. 

Since a smaller ICOBI indicates a more ionic bonding character, this trend suggests that the TM--O$^*$ bonds become increasingly ionic as the 3$d$ orbitals are progressively filled. In 3$d$ TMO$_2$, an increased $d$-electron count is often associated with reduced $d$--$p$ hybridization and enhanced localization of the 3$d$ orbitals, which can favor ionic bonding character~\cite{3d-2pHybridLocal}. However, this tendency is not universal and depends sensitively on factors such as the relative $d$--$p$ energy alignment and oxidation state. Nevertheless, stronger ionic character is typically accompanied by a reduction in the effective ionic radius of the TM ion, which in turn enhances the electrostatic attraction between TM$^{4+}$ and O$^{2-}$. This interpretation is consistent with the observed trend in Fig.~\ref{FIG_fit_2}(a).

\subsection{Migration barrier}
\label{sec:migration}


The DFT-cNEB energy profile ($E^{\rm DFT}$) of migrating $V_{\rm O}$ are compared with $S_{c}$ and $S_{i}$ (Fig.~\ref{FIG_barrier}). Since the ICOHP is defined in units of energy, its values can be compared directly with $E^{\rm DFT}$. Overall, although $S_{c}$ and $S_{i}$ are not perfectly matched with $E^{\rm DFT}$, their maximum values are intimately connected to $E_{\rm B}^{\rm DFT}$. For TiO$_2$, our explicit DFT-cNEB calculation gives $E_{\rm B}^{\rm DFT}$ of 1.25 eV, which is close to the other DFT-GGA result of 1.11 eV~\cite{Zhu_2014} and to the experimental value of 1.15 eV~\cite{TiO2_exp_barrier}. But, it is much smaller than the DFT-HSE value of 1.52 eV~\cite{Zhu_2014}. The maximum values of $S_{c}$ ($S_{\rm c, max}$), $S_{i}^{L}$ ($S_{i,\rm {max}}^{L}$), and $S_{i}^{M}$ ($S_{i,\rm {max}}^{M}$) are 1.48 eV, 0.88 eV, and 1.46 eV, respectively. For VO$_2$, we find $E_{\rm B}^{\rm DFT}$ of 0.96 eV, which is close to another GGA+$U$ value of 0.69 eV~\cite{sim2024crystallographic}. $S_{\rm c, max}$, $S_{i,\rm {max}}^{L}$, and $S_{i,\rm {max}}^{M}$ are 1.04 eV, 1.33 eV, and 1.92 eV, respectively.

\begin{figure}[b]
	\includegraphics[width= 8 cm]{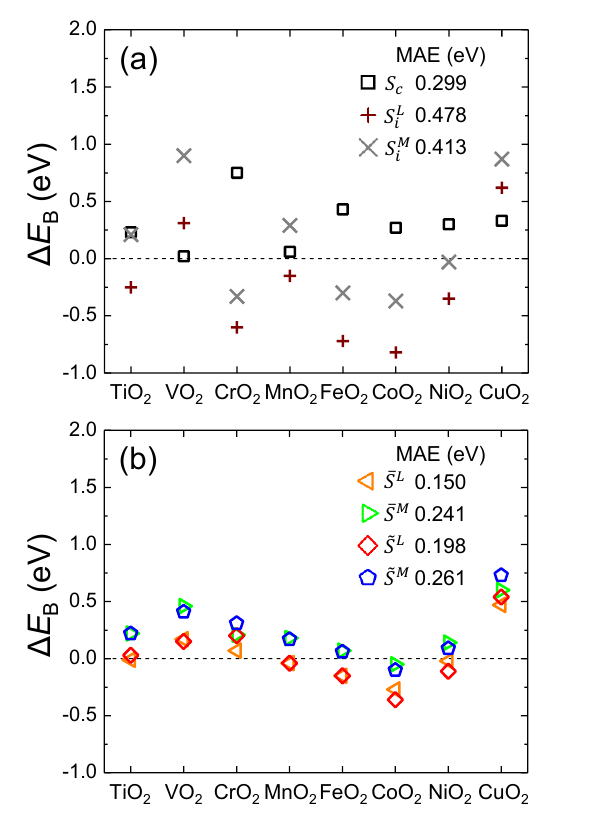}\\
	\caption{ \label{FIG_ave} (a) Energy difference ($\Delta E_{\rm B}$) from $E_{\rm B}^{\rm DFT}$. (a) $S_{\rm c, max}$, $S_{i,\rm {max}}^{L}$, and $S_{i,\rm {max}}^{M}$. (b) $\bar{S}^{L}$, $\bar{S}^{M}$, $\tilde{S}^{L}$, and $\tilde{S}^{M}$. 
    MAE denotes mean absolute error.}
        
\end{figure}

Note that the migration path of $V_{\rm O}$ reported in the literature are not identical. For example, Zhu {\it et al.}~\cite{Zhu_2014} considered three paths in the rutile TiO$_2$ and the path involving the lowest barrier depends on computational details such as supercell size and $k$-point mesh. Thus, we selected a value in the literature that corresponds to the migration path considered in the present work for a fair comparison.

\begin{figure*}[]
	\includegraphics[width=2\columnwidth]{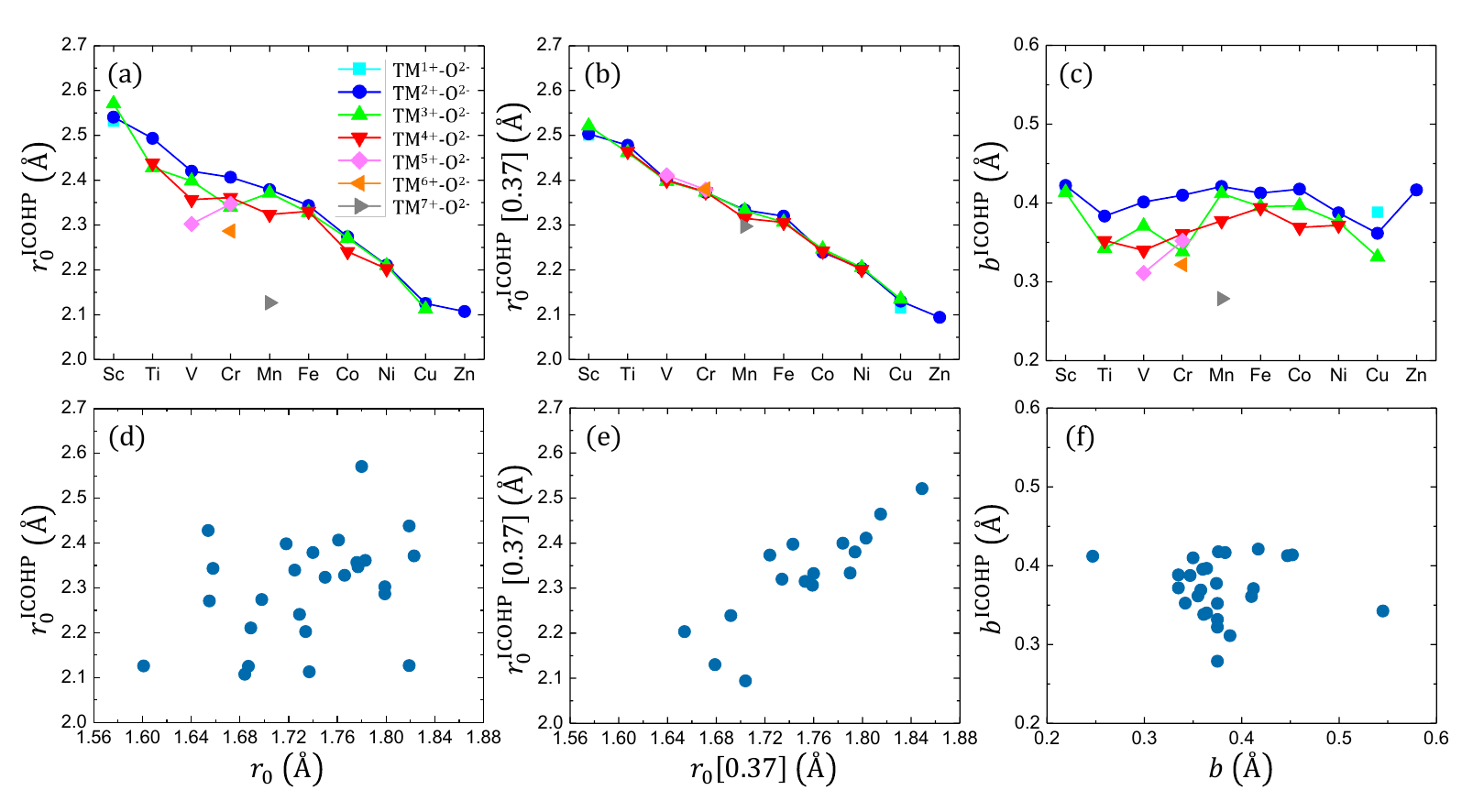}\\
	\caption{ \label{FIG_fit_1} (a) $r_0^{\rm ICOHP}$, (b) $r_0^{\rm ICOHP}$[0.37], and (c) $b^{\rm ICOHP}$ for TM--O$^*$ across 3$d$ TM element from Sc to Zn. (d)--(f) Comparison between the ICOHP-based BVM parameters and the BVM parameters. 
	}
\end{figure*}

Although $E_{\rm B}^{\rm DFT}$ appears to be related to the bond strength quantities, the closest quantity seems to vary depending on the material. For some materials, $S_{\rm c, max}$ has the closest value, while for others, $S_{i,\rm {max}}^{L}$ or $S_{i,\rm {max}}^{M}$ is closer than $S_{\rm c, max}$. For CoO$_2$, $S_{\rm c, max}$ (1.90 eV) is comparable to and much closer to $E_{\rm B}^{\rm DFT}$ (1.63 eV) than $S_{i,\rm {max}}^{L}$ (0.81 eV) and $S_{i,\rm {max}}^{M}$ (1.26 eV). For CrO$_2$, $S_{i,\rm {max}}^{M}$ (0.84 eV) is closer to $E_{\rm B}^{\rm DFT}$ (1.17 eV) than $S_{\rm c, max}$(1.92 eV) and $S_{i,\rm {max}}^{L}$ (0.57 eV). 

As seen in Fig.~\ref{FIG_ave}(a), the energy difference between the bond strength quantities and $E_{\rm B}^{\rm DFT}$ ($\Delta E_{\rm B}$) is plotted by defining as $\alpha - E_{\rm B}^{\rm DFT}$ ($\alpha$ = $S_{\rm c, max}$, $S_{i,\rm {max}}^{L}$, and $S_{i,\rm {max}}^{M}$). There exists sizable deviations from $E_{\rm B}^{\rm DFT}$ with the mean absolute error (MAE) of 0.260 eV to 0.475 eV. I.e., $S_{\rm c, max}$ works better than $S_{i,\rm {max}}$ overall. 

In general, materials exhibit mixed bonding characteristics, and oxides in particular are largely characterized by a combination of covalent and ionic bonding. Accordingly, the representative value was evaluated by averaging those of $S_{\rm c}$ and $S_{i}$; 
\begin{eqnarray} \label{eqave1}
\bar{S}^{L} = (S_{\rm c, max} + S_{i,\rm {max}}^{L})/2\\
\bar{S}^{M} = (S_{\rm c, max} + S_{i,\rm {max}}^{M})/2
\end{eqnarray}
Figure~\ref{FIG_ave}(b) and Table~\ref{barrier1} clearly show that $\bar{S}^{L}$ and $\bar{S}^{M}$ are found to be comparable to $E_{\rm B}^{\rm DFT}$ (i.e.,$\alpha$ = $\bar{S}^{L}$ and $\bar{S}^{M}$). For TiO$_2$, $\bar{S}^{L}$ and $\bar{S}^{M}$ are, respectively, 1.24 eV and 1.47 eV, which excellently agree with $E_{\rm B}^{\rm DFT}$ of 1.25 eV. CrO$_2$, in which $S_{\rm c}$ and $S_{i}$ are far from $E_{\rm B}^{\rm DFT}$, has 1.38 eV for $\bar{S}^{M}$ and 1.25 eV for $\bar{S}^{L}$. Indeed, these are close to $E_{\rm B}^{\rm DFT}$ of 1.17 eV. CoO$_2$ has 1.58 eV for $\bar{S}^{M}$ and 1.36 eV for $\bar{S}^{L}$, which are also close to the $E_{\rm B}^{\rm DFT}$ of 1.63 eV.




The ICOBI-based weighted average is further examined, as follows: 
\begin{eqnarray} \label{eqave2}
\tilde{S}^{L} = {\rm ICOBI} \times S_{\rm c, max}+ (1 - {\rm ICOBI}) \times S_i^{L}\\
\tilde{S}^{M} = {\rm ICOBI} \times S_{\rm c, max} + (1 - {\rm ICOBI}) \times S_i^{M}
\end{eqnarray}
where ICOBI is the value for each material as seen in Fig.~\ref{FIG_fit_2}(a). Fig.~\ref{FIG_ave}(b) shows that the ICOBI-based weighting (i.e.,$\alpha$ = $\tilde{S}^{L}$ and $\tilde{S}^{M}$) does not improve the accuracy. The MAE for $\tilde{S}^{L}$ and $\tilde{S}^{M}$ is slightly higher than that of $\bar{S}^{L}$ and $\bar{S}^{M}$.


\subsection{ICOHP-based BVM parameters}
\label{sec:comparison}

\begin{table*}[]
\centering
    \caption{\label{barrier} Fitted $r_{0}^{\rm ICOHP}$ and $b^{\rm ICOHP}$ for $\rm TM^{4+}$--$\rm O^{2-}$, $\rm TM^{4+}$--$\rm TM^{4+}$, and $\rm O^{2-}$--$\rm O^{2-}$. $r_{0}$ and $b$ in the BVM are shown for comparison. $r_{0}^{\mathrm{ICOHP}}[0.37]$ and $r_{0}[0.37]$ correspond to the values obtained with $b_{\mathrm{ICOHP}}$ = $b$ = 0.37 \AA.}
    \label{TAB:params}
    \begin{ruledtabular}
        \begin{tabular}{c cc cc cc}
            TM
            & $r_{0}^{\rm ICOHP}$ (\AA)
            & $r_{0}$~\cite{bvpara2} (\AA)
            & $b^{\rm ICOHP}$ (\AA)
            & $b$~\cite{bvpara2} (\AA)
            & $r_{0}^{\rm ICOHP}[0.37]$ (\AA)
            & $r_{0}[0.37]$~\cite{bvpara1} (\AA) \\
           \hline
			$\rm Ti^{4+}$--$\rm O^{2-}$ & 2.438 & 1.819 & 0.352 & 0.342 & 2.464 & 1.815 \\
            $\rm V^{4+}$--$\rm O^{2-}$ & 2.357 & 1.776 & 0.340 & 0.364 & 2.400 & 1.784 \\
            $\rm Cr^{4+}$--$\rm O^{2-}$ & 2.361 & 1.783 & 0.361 & 0.410 & 2.374 & \\
            $\rm Mn^{4+}$--$\rm O^{2-}$ & 2.324 & 1.750 & 0.377 & 0.374 & 2.315 & 1.753 \\
            $\rm Fe^{4+}$--$\rm O^{2-}$ & 2.331 &  & 0.394 &  & 2.306 &  \\
            $\rm Co^{4+}$--$\rm O^{2-}$ & 2.241 & 1.729 & 0.369 & 0.358 & 2.242 &  \\
            $\rm Ni^{4+}$--$\rm O^{2-}$ & 2.202 & 1.734 & 0.372 & 0.335 & 2.201 &  \\
			$\rm Ti^{4+}$--$\rm Ti^{4+}$ & 2.795 &  & 0.438 &  & 2.833 &  \\
            $\rm V^{4+}$--$\rm V^{4+}$ & 2.727 &  & 0.417 &  & 2.744 &  \\
            $\rm Cr^{4+}$--$\rm Cr^{4+}$ & 2.404 &  & 0.323 &  & 2.323 &  \\
            $\rm Mn^{4+}$--$\rm Mn^{4+}$ & 1.870 &  & 0.531 &  & 2.176 &  \\
            $\rm Fe^{4+}$--$\rm Fe^{4+}$ & 2.114 &  & 0.361 &  & 2.091 &  \\
            $\rm Co^{4+}$--$\rm Co^{4+}$ & 2.777 &  & 0.091 &  & 2.558 &  \\
            $\rm Ni^{4+}$--$\rm Ni^{4+}$ & 1.491 &  & 0.580 &  & 1.979 &  \\
            $\rm O^{2-}$--$\rm O^{2-}$ & 2.030 &  & 0.276 &  & 1.967 &  \\
            \end{tabular}
    \end{ruledtabular}
\end{table*}

As mentioned earlier, the ICOHP is a quantitative measure of covalent bond strength based on orbital interactions, analogous to the BVM. Thus, the ICOHP can be expressed in a form similar to that of the BVM ( Eq.~(\ref{eq1})), as follows, 
\begin{eqnarray} \label{eq6}
	-\mathrm{ICOHP} = {\rm exp}\left (- \frac{r-r_{0}^{\rm ICOHP}}{b^{\rm ICOHP}}\right)
\end{eqnarray}
where $r$ is the bond length.  $r_0^{\rm ICOHP}$ and $b^{\rm ICOHP}$ are bond strength parameters that can be obtained by fitting. A prefactor of 1 eV is assumed in front of the exponential term.

This exponential functional form can be justified by the quantum mechanical nature of orbital interactions. In the linear combination of atomic orbitals framework, the covalent interaction energy is intrinsically proportional to the overlap integral between adjacent atomic orbitals. Because the radial part of localized atomic orbitals, such as Slater-type orbitals, decays exponentially at large interatomic distances ($ \propto \exp(-\zeta r)$ where $\zeta$ represents the Slater orbital exponent, which dictates the spatial extent and the radial decay rate of the atomic electron cloud)~\cite{slater1930atomic}, their corresponding overlap integrals also exhibit an exponential decay~\cite{burdett1993orbital}. Consequently, the ICOHP, which quantifies the covalent bond strength derived from these overlapping electronic states, may follow an exponential dependence on the interatomic distance $r$. 


The ICOHP-based BVM parameters $r_{0}^{\rm ICOHP}$ and $b^{\rm ICOHP}$ are obtained using the 795 data set collected from the {\sc Materials Project} database ~\cite{mp}, comprising oxides formed with 3$d$ TM (e.g., TM$_2$O$_3$ and TM$_3$O$_4$) that span Sc to Zn, covering a wide range of local bonding environments and oxidation states. Specifically, the number of structures included for each element is 13 for Sc, 124 for Ti, 209 for V, 90 for Cr, 67 for Mn, 140 for Fe, 60 for Co, 36 for Ni, 38 for Cu, and 18 for Zn. These structures serve as the reference data set for the parameterization. The results are summarized in Table~\ref{TAB:params}.

Overall, $b^{\mathrm{ICOHP}}$ and $b$ are similar, whereas $r_{0}^{\rm ICOHP}$ and $r_{0}$ differ significantly. For $\rm Ti^{4+}$--$\rm O^{2-}$, $b^{\rm ICOHP}$ of 0.352 {\AA} is comparable to the BVM parameter $b$ of 0.342 {\AA} reported in the BVM~\cite{bvpara2}. On the other hand, $r_{0}^{\rm ICOHP}$ of 2.438 {\AA} differs from the BVM value of $r_0$ = 1.819 {\AA}. 
For $\rm Mn^{4+}$--$\rm O^{2-}$, $b^{\rm ICOHP}$ of 0.377 {\AA} and $b$ of 0.374 {\AA} are close to each other, while $r_{0}^{\rm ICOHP}$ of 2.324 {\AA} quite differs from $r_0$ of 1.750 {\AA}. These may reflect the distinct physical basis of the two approaches. Unlike the BVM, which relies on predefined oxidation states and empirical distance--valence relationships, our approach is applicable to interactions between identical species (e.g. TM--TM, O--O) and to systems in which the ions share the same formal oxidation state (e.g. TM$^{4+}$--TM$^{4+}$, O$^{2-}$--O$^{2-}$).

We also find that the parameters are insensitive to  change in the TM oxidation state. For example, $r_{0}^{\rm ICOHP}$ of Ti$^{3+}$-- O$^{2-}$ is 2.428 {\AA}, which is slightly smaller than that of Ti$^{4+}$-- O$^{2-}$ (2.438 {\AA}). This arises from the fact that $r_{0}^{\rm ICOHP}$ and $b^{\rm ICOHP}$ are derived directly from orbital-resolved electronic-structure information rather than from fixed ionic assumptions. Accordingly, the ICOHP-based parameterization might provide a systematic and transferable framework for extracting $r_{0}^{\rm ICOHP}$ and $b^{\rm ICOHP}$ across diverse chemical environments.

Figures~\ref{FIG_fit_1}(a) and (b) present the calculated $r_0^{\rm ICOHP}$ and $r_0^{\rm ICOHP}$[0.37] for each TM--O in TMO$_2$. $r_0^{\rm ICOHP}$ linearly decreases with increasing 3$d$ electron count of the TM elements. This becomes more clear when $b^{\rm ICOHP}$ is fixed at $b$ (=0.37 {\AA}). The function $r_0^{\rm ICOHP}$[0.37] = --0.0452 (3$d$ electron count) + 2.5535 (R$^2$ = 0.9832) is obtained by linear fit. In contrast, no clear trend is observed for $b^{\rm ICOHP}$ (Fig.~\ref{FIG_fit_1}(c)).
A comparison between $r_{0}^{\rm ICOHP}$ and $b^{\rm ICOHP}$ and the BVM parameters is illustrated to examine their relationship. No clear correlation with the BVM parameters is observed for $r_0^{\rm ICOHP}$ vs $r_0$ (Fig.~\ref{FIG_fit_1}(d)) and for $b^{\rm ICOHP}$ vs $b$ (Fig.\ref{FIG_fit_1}(f)). On the other hand, a linear-like tendency is seen for $r_0^{\rm ICOHP}$[0.37] vs $r_0$[0.37] (Fig.~\ref{FIG_fit_1}(e)).


We now examine the physical meaning of the ICOHP-based parameters. As shown in Fig.~\ref{FIG_fit_2}(b), $r_{0}^{\mathrm{ICOHP}}$ exhibits a linear correlation with the sum of ionic radii of TM$^{4+}$ and O$^{2-}$ ions, indicating that, analogous to the BVM, $r_{0}^{\mathrm{ICOHP}}$ retains its interpretation as a nominal bond length. In contrast, $b^{\mathrm{ICOHP}}$ shows a distinct trend compared to the conventional BVM parameter $b$. Previous studies based on the BVM reported a nonlinear dependence of $b$ on the softness difference $(\sigma_{\rm anion} - \sigma_{\rm cation})$, where $b$ increases as the magnitude of the softness mismatch becomes larger~\cite{adams2001relationship,Adams:br0103}. The slope of $b$ changes sign between the positive and negative regimes of $(\sigma_{\rm anion} - \sigma_{\rm cation})$, reflecting the approximately U-shaped nonlinearity of the empirical relationship. 

In rutile TMO$_2$ considered here, we found that $(\sigma_{\rm O} - \sigma_{\rm TM})$ lies entirely in the negative regime, where $\sigma_{\rm TM} > \sigma_{\rm O}$. Within this restricted range, $b^{\mathrm{ICOHP}}$ increases with $(\sigma_{\rm O} - \sigma_{\rm TM})$, corresponding to an apparent positive slope with respect to $(\sigma_{\rm O} - \sigma_{\rm TM})$ (Fig.~\ref{FIG_fit_2}(c)). This behavior differs from the negative slope observed in the corresponding regime of the empirical BVM relation~\cite{adams2001relationship,Adams:br0103}.

This difference originates from the distinct physical basis of the two approaches. While the BVM parameter $b$ reflects an empirical description of bond valence within an ionic framework based on the softness mismatch between ions, $b^{\mathrm{ICOHP}}$ is directly derived from orbital interactions. Chemical softness represents the polarizability and spatial extent of the atomic electron cloud~\cite{geerlings2003conceptual}, such that softer elements exhibit more delocalized orbitals. This leads to a slower decay of orbital overlap with interatomic distance. Therefore, $b^{\mathrm{ICOHP}}$ can be interpreted as a parameter describing the spatial decay length of orbital interactions. As a result, in systems with significant covalent character, such as rutile TMO$_2$~\cite{Soulard_2003,Toropova_2005}, $b_{\mathrm{ICOHP}}$ is primarily governed by the degree of orbital delocalization rather than the ionic softness mismatch. This may explain why the trend of $b_{\mathrm{ICOHP}}$ with respect to softness differs from that of the empirical BVM parameter in the negative softness-difference regime.

\subsection{Estimation of migration barrier}
\label{sec:prediction}

\begin{figure}[t]
    \includegraphics[width=7.5 cm]{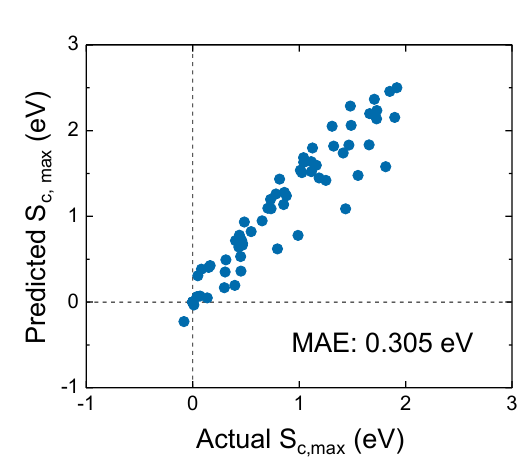}\\
	\caption{ \label{FIG_SC} Comparison between $S_{\rm c, max}$ evaluated using Eq.~(\ref{eq6}) with $r_0^{\rm ICOHP}$ and $b^{\rm ICOHP}$ (Predicted) and obtained by the ICOHP (Actual values). The results are obtained for the structures corresponding to the images along the migration pathway in the DFT-cNEB calculation.
}
\end{figure}

\begin{figure}[t]
		\includegraphics[width=8.5 cm]{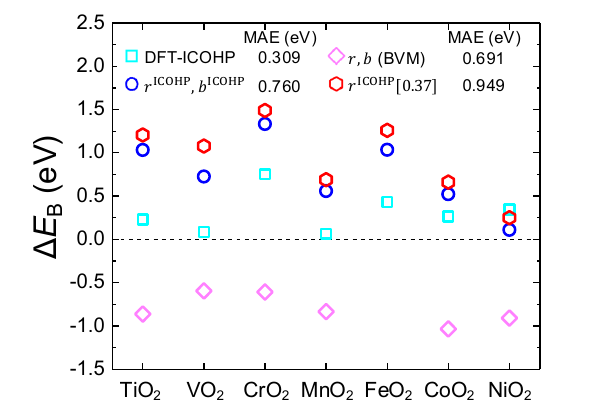}\\
	\caption{ \label{FIG_prediction} $\Delta E_{\rm B}$ with $S_{\rm c, max}$ from DFT-ICOHP, the ICOHP-based BVM parameters ($r_0^{\rm ICOHP}$ and $b^{\rm ICOHP}$), the BVM parameters ($r_0$ and $b$), and $r_0^{\rm ICOHP}[0.37]$.}
\end{figure}

Figure~\ref{FIG_SC} compares $S_{\rm c,max}$ obtained from DFT-ICOHP calculations ("Actual") with those estimated from --ICOHP using Eq.~(\ref{eq6}) ("Predicted"). The predicted $S_{\rm c,max}$ shows a clear linear correlation with the actual values obtained from explicit ICOHP calculations. This indicates that, once $r_0^{\rm ICOHP}$ and $b^{\rm ICOHP}$ are determined, $E_{\rm B}$ can be approximately estimated without performing explicit DFT-cNEB calculations, as $S_{\rm c,max}$ is found to be reasonably close to $E_{\rm B}^{\rm DFT}$ (Fig.~\ref{FIG_barrier} and Fig.~\ref{FIG_ave}(a)).

Therefore, we further estimate $E_{\rm B}$ using $S_{\rm c,max}$ derived from $r_0^{\rm ICOHP}$ and $b^{\rm ICOHP}$, and compare the results with those obtained from DFT-ICOHP, the BVM parameters ($r_0$ and $b$), and $r_0^{\rm ICOHP}[0.37]$. As shown in Fig.~\ref{FIG_prediction}, $S_{\rm c,max}$ based on $r_0^{\rm ICOHP}$ and $b^{\rm ICOHP}$ tends to overestimate $E_{\rm B}$, whereas that based on the BVM parameters ($r_0$ and $b$) underestimates it. However, both approaches yield comparable absolute errors in $\Delta E_{\rm B}$, with similar MAEs. Interestingly, the estimation using $r_0^{\rm ICOHP}[0.37]$ performs the worst, which is not expected from the trend observed in Fig.~\ref{FIG_fit_1}(b). Note that unlike Fig.~\ref{FIG_ave}, Fig.~\ref{FIG_prediction} does not include data for CuO$_2$. The BVM does not provide Cu$^{4+}$--O$^{2-}$ pairs for analysis, and therefore the corresponding $b^{\mathrm{ICOHP}}$ and $r_{0}^{\mathrm{ICOHP}}$ parameters could not be determined.

Although $S_{\rm c,max}$ obtained from $r_0^{\rm ICOHP}$ and $b^{\rm ICOHP}$ is in reasonable agreement with $E_{\rm B}^{\rm DFT}$ for certain cases such as NiO$_2$ and CoO$_2$, the overall MAE remains relatively large. This discrepancy can be attributed to the lack of an explicit description of ionic contributions in Eq.~(\ref{eq6}) and the absence of a complete treatment of the mixed ionic–covalent nature of bonding. This observation supports the conclusion that a combined analysis of ionic and covalent aspects of bonding is essential, governed by electronic structure factors such as hybridization and orbital localization, play important roles in determining the bond strength and its distance dependence in TMO$_2$. Nevertheless, since conventional COHP calculations require prior DFT electronic-structure calculations, the present approach offers a significant computational advantage by enabling rapid evaluation of bond strength without explicitly resolving the full electronic structure for every single system.

\section{Conclusion}

Using DFT and the COHP method, we investigated $E_B$ of $V_{\mathrm{O}}$ in rutile-type 3$d$ TMO$_2$ by establishing a direct connection between bond strength and ion migration. The bond strength associated with the migrating oxygen atom was decomposed into covalent and ionic contributions, quantified by $S_c$ and $S_i$, respectively. Both $S_c$ and $S_i$ correlate with $E_B$ from DFT calculation, i.e., $E_B^{\mathrm{DFT}}$, but neither alone provides a universally accurate descriptor across different materials. Instead, their simple average yields a robust and consistent estimate of $E_B^{\mathrm{DFT}}$ over the entire series, highlighting the cooperative nature of mixed covalent–ionic bonding in determining migration energetics.

Motivated by the BVM, we further introduced an ICOHP-based parameterization in which the covalent bond strength is expressed as an exponential function of interatomic distance. From a large dataset of TM oxides, two parameters, $r_0^{\mathrm{ICOHP}}$ and $b^{\mathrm{ICOHP}}$, were extracted. These parameters provide a physically transparent description of bonding directly derived from electronic structure, without relying on predefined oxidation states or empirical ionic assumptions. The proposed framework enables an efficient estimation of $E_B$ without performing computationally demanding DFT calculations. Despite its simplicity, the obtained $S_c$ based on the parameters reasonably predict $E_B^{\mathrm{DFT}}$, which indicates that the approach captures the essential physics of bond breaking and reformation along the migration pathway.

\begin{acknowledgments}


This work was supported by Inha University Research Grant (INHA-74153). This research
used resources of the National Supercomputing Center with supercomputing resources including technical support (Grant No. KSC-2024-CRE-0395, KSC-2025-CRE-0596).

\end{acknowledgments}

\bibliography{migration}

\end{document}